\documentclass[lettersize,journal]{IEEEtran}
\usepackage{amsmath,amsfonts}
\usepackage{algorithmic}
\usepackage{algorithm}
\usepackage{array}
\usepackage[caption=false,font=normalsize,labelfont=sf,textfont=sf]{subfig}
\usepackage{textcomp}
\usepackage{stfloats}
\usepackage{url}
\usepackage{verbatim}
\usepackage{graphicx}
\usepackage{cite}
\usepackage{xcolor}
\def\BibTeX{{\rm B\kern-.05em{\sc i\kern-.025em b}\kern-.08em
    T\kern-.1667em\lower.7ex\hbox{E}\kern-.125emX}}
\begin{document}

\title{Timing Recovery for Non-Orthogonal Multiple Access with Asynchronous Clocks}

\author{Qingxin Lu, Haide Wang, Wenxuan Mo, Ji Zhou, Weiping Liu, and Changyuan Yu

\thanks{Manuscript received 3 June 2024; revised 17 August 2024; accepted 6 September 2024. This work is supported in part by the National Key R$\&$D Program of China under Grant 2023YFB2905700, in part by the National Natural Science Foundation of China under Grant 62371207 and Grant 62005102, in part by the Young Elite Scientists Sponsorship Program by CAST under Grant 2023QNRC001, and in part by the Hong Kong Research Grants Council GRF under Grant 15231923. \it{(Corresponding author: Ji Zhou.)}}
\thanks{Qingxin Lu, Ji Zhou, and Weiping Liu are with the Department of Electronic Engineering, College of Information Science and Technology, Jinan University, Guangzhou 510632, China.}
\thanks{Haide Wang is with the School of Cyber Security, Guangdong Polytechnic Normal University, Guangzhou 510665, China.}
\thanks{Wenxuan Mo is with the Guangdong Planning and Designing Institute of Telecommunications Co., Ltd., Guangzhou 510630, China.}
\thanks{Changyuan Yu is with the Department of Electrical and Electronic Engineering, The Hong Kong Polytechnic University, Hong Kong.}}

\markboth{IEEE Photonics Technology Letters}%
{Shell \MakeLowercase{\textit{et al.}}: A Sample Article Using IEEEtran.cls for IEEE Journals}


\maketitle

\begin{abstract} 
A passive optical network (PON) based on non-orthogonal multiple access (NOMA) meets low latency and high capacity. In the NOMA-PON, the asynchronous clocks between the strong and weak optical network units (ONUs) cause the timing error and phase noise on the signal of the weak ONU. The theoretical derivation shows that the timing error and phase noise can be independently compensated. In this Letter, we propose a timing recovery (TR) algorithm based on an absolute timing error detector (Abs TED) and a pilot-based carrier phase recovery (CPR) to eliminate the timing error and phase noise separately. An experiment for 25G NOMA-PON is set up to verify the feasibility of the proposed algorithms. The weak ONU can achieve the 20\% soft-decision forward error correction limit after compensating for timing error and phase noise. In conclusion, the proposed TR and the pilot-based CPR show great potential for the NOMA-PON.
\end{abstract}


\section{Introduction}\label{Introduction}
Passive optical network (PON) with a point-to-multi-point architecture is widely regarded as a promising solution for not only access networks for homes and base stations but also some short-reach access scenarios, such as networks in vehicles or rooms \cite{wu2023fiber, van2020strategies, liu2023optical}. Except for wavelength division multiple access, time division multiple access, and frequency division multiple access, non-orthogonal multiple access (NOMA) can also be used in PON. In NOMA-PON, the signals from strong and weak optical network units (ONUs) are multiplexed in the power domain to meet low latency and high capacity \cite{mo2022adaptive, sarmiento2020optical, liaqat2020power, lin2017experimental}. To recover the signals of the strong and weak ONUs, successive interference cancellation (SIC) has been widely studied, including SIC for synchronized and asynchronous NOMA \cite{lu2017power, lin2019power}, SIC with carrier frequency offset compensator \cite{suzuoki2021demonstration}, and SIC with dynamic power control \cite{sun2021new}. 

In PON, the clock frequency of ONU can be locked based on that extracted from the downstream signal of the optical line terminal (OLT). However, the clock of the strong and weak ONUs in the NOMA-PON is asynchronous, which inevitably causes a timing jitter between the upstream signals from the strong and weak ONUs. Therefore, the timing recovery (TR) algorithm is required to correct the timing error on the signal of weak ONU caused by the asynchronous clocks. In the signal of weak ONU at baud-rate sampling, the TR algorithm commonly uses the Mueller-Muller timing error detector (TED) \cite{mueller1976timing, musah2022robust}. However, in NOMA-PON with the asynchronous clocks, the frequency shift and timing error result in phase noise, which makes the Mueller-Muller TED based on the amplitude variation trend ineffective. To the best of our knowledge, most of the relative works focus on the SIC, while there is almost no research on the TR for NOMA with asynchronous clocks. Thus, a more suitable TR algorithm and carrier phase recovery (CPR) should be proposed for NOMA-PON.

In this Letter, the timing error and phase noise on the signal of weak ONU caused by the asynchronous clocks are theoretically derived, which shows that the timing error and phase noise can be independently compensated. The TR algorithm with absolute (Abs) TED is proposed to eliminate the timing error, which is insensitive to phase noise. Meanwhile, pilot-based CPR is proposed to reduce phase noise. The experimental results show that the TR with Abs TED and pilot-based CPR can effectively compensate for the timing error and phase noise caused by the asynchronous clocks.

\section{Principle of NOMA with Asynchronous Clocks}\label{Principle}
In this section, a detailed theoretical derivation for the generation of phase noise
in NOMA-PON is given. The signal of strong ONU after frequency shift can be expressed as
\begin{equation}
x_{1}(t) = I_{1}(t)\cdot \cos (2\pi f_{c} t) + Q_{1}(t)\cdot \sin(2\pi f_{c} t)
\label{Eq1}
\end{equation}
where $I_{1}(t)$ and $Q_{1}(t)$ represent the in-phase (I) and quadrature (Q) tributaries of the signal of strong ONU, respectively. $f_{c}$ is the central frequency of the digital subcarrier. Due to the asynchronous clocks, the signal of weak ONU with timing jitter $\tau$ can be represented as
\begin{equation}
\begin{aligned}
x_{2}(t+\tau) & = I_{2}(t+\tau)\cdot \cos\left[2\pi f_{c} (t+\tau)\right]\\
&+Q_{2}(t+\tau)\cdot\sin\left[2\pi f_{c} (t+\tau)\right]
\label{Eq2}
\end{aligned}
\end{equation}
where $I_2(t+\tau)$ and $Q_2(t+\tau)$ are the I and Q tributary signals of weak ONU, respectively. In NOMA-PON, the strong and weak ONUs have the same center frequency. The signal $x_{1}(t)$ of strong ONU and the signal $x_{2}(t)$ of weak ONU multiplexed in the optical power domain to generate the NOMA signal.

The optical field of the NOMA signal can be expressed as
\begin{equation}
\begin{aligned}
x_{\text{NOMA,~O}}(t)={}& A_{1}\left[1+\alpha_{1} x_{1}(t)\right]e^{j2\pi f_{0} t}+\\
& A_{2}\left[1+\alpha_{2} x_{2}(t+\tau +\Delta t)\right]e^{j2\pi (f_{0}+\Delta f)t}
\label{Eq3}
\end{aligned}
\end{equation}
where $\Delta t$ and $\Delta f$ are the transmission delay and the frequency difference between the two lasers of the ONUs, respectively. $A_{1}$, $A_{2}$, $\alpha_{1}$, $\alpha_{2}$ are the amplitude, modulation index of strong ONU and weak ONU, respectively. Generally, $\alpha_{1}=\alpha_{2}=\alpha$. The power ratio in dB can be calculated by $20 \log_{10} (A_{1} / A_{2})$. $f_{0}$ is the central frequency of the optical signal.

\begin{figure}[!t]
\centering
\includegraphics[width = 0.9\linewidth]{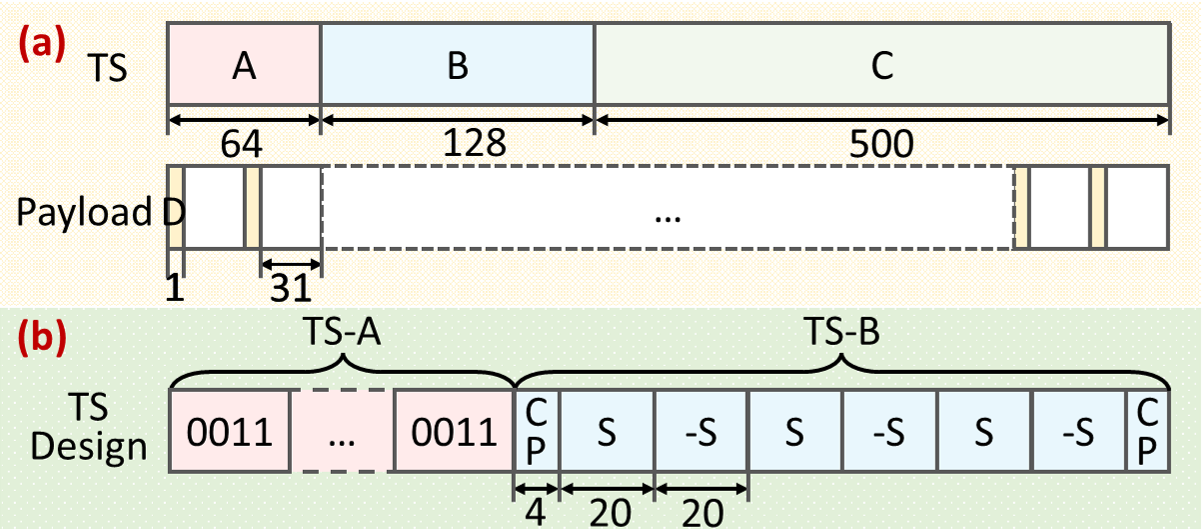}
\caption{(a) The frame structure and training sequence (TS) design for the NOMA signals. (b) The structure design of TS-A and TS-B.}
\label{Fig1}
\end{figure}

After the optical-to-electrical square-law detection, the electrical signal $y(t)$ can be expressed as
\begin{equation}
\begin{aligned}
 y(t) & \propto x_{\text{NOMA,~O}}(t) \times \left[x_{\text{NOMA,~O}}(t)\right]^\ast\\
&=A_1^2\left[1+\alpha x_{1}(t)\right]^2+A_2^2\left[1+\alpha x_{2}(t+\tau +\Delta t)\right]^2+\\
&2A_1A_2\left[1+\alpha x_{1}(t)\right]\left[1+\alpha x_{2}(t+\tau +\Delta t)\right]\cos(\Delta f t)
\label{Eq4}
\end{aligned}
\end{equation}
where $(\cdot)^\ast$ denotes a conjugate operation, $x_{1}(t)$ and $x_{2}(t)$ fulfill $\lvert\alpha x_{1}(t)\rvert \ll 1$ and $\lvert\alpha x_{2}(t+\tau+\Delta t)\rvert \ll 1$, respectively \cite{xu2016bidirectional}. Due to the limited bandwidth, the third term of Eq. (\ref{Eq4}) will be filtered out and the direct current (DC) term is no longer considered. Therefore, when the modulator is operated under a linear region, the received signal can be expressed as
\begin{equation}
\begin{aligned}
y(t)\propto A_1^2 x_{1}(t) + A_2^2 x_{2}(t+\tau+\Delta t).
\label{Eq5}
\end{aligned}
\end{equation}
The baseband NOMA signal $y_{\text{NOMA}}(t)$ is obtained from $y(t)$ after the frequency shift. Since the high-frequency signal is filtered out, the $y_{\mathrm{NOMA}}(t)$ can be represented as
\begin{equation}
\begin{aligned}
y_{\text{NOMA}}(t) & \propto A_1^2 \left[I_{1}(t)+j Q_{1}(t)\right]+ A_2^2 \left[I_{2}(t+\tau+\Delta t)\right.\\
&\left.-j Q_{2}(t+\tau+\Delta t)\right] e^{-j 2 \pi f_{c} (\tau+\Delta t)}
\label{Eq6}
\end{aligned}
\end{equation}
where $e^{-j2 \pi f_{c} (\tau+\Delta t)}$ is the phase noise caused by the frequency shift. The timing error and phase noise can be independently compensated because the timing information $t$ is not concluded in the phase noise. Depending on the power of the users, the strong ONU signal $y_1(t)$ is obtained by $y_{\text{NOMA}}(t)$ after a hard decision using SIC. The signal $y_2(t)$ of weak ONU can be calculated by $y_{\text{NOMA}}(t) - y_1(t)$. The frame synchronization can recover the $\Delta t$. 

At the baud-rate sampling, the TR with Abs TED uses the absolute values of the samples to estimate the timing error, which is insensitive to the phase noise \cite{stojanovic2020baud}. The timing error $E_{\text{Abs}}(n)$ can be estimated as
\begin{equation}
\begin{aligned}
E_{\text{Abs}}(n) = \lvert y_2(n-1)+y_2(n)\rvert \cdot \left[\lvert y_2(n-1)\rvert-\lvert y_2(n)\rvert\right]
\label{Eq7}
\end{aligned}
\end{equation}
where $y_2(n)$ is the $n$-th symbol of weak ONU.

\begin{figure}[!t]
\centering
\includegraphics[width = 0.9\linewidth]{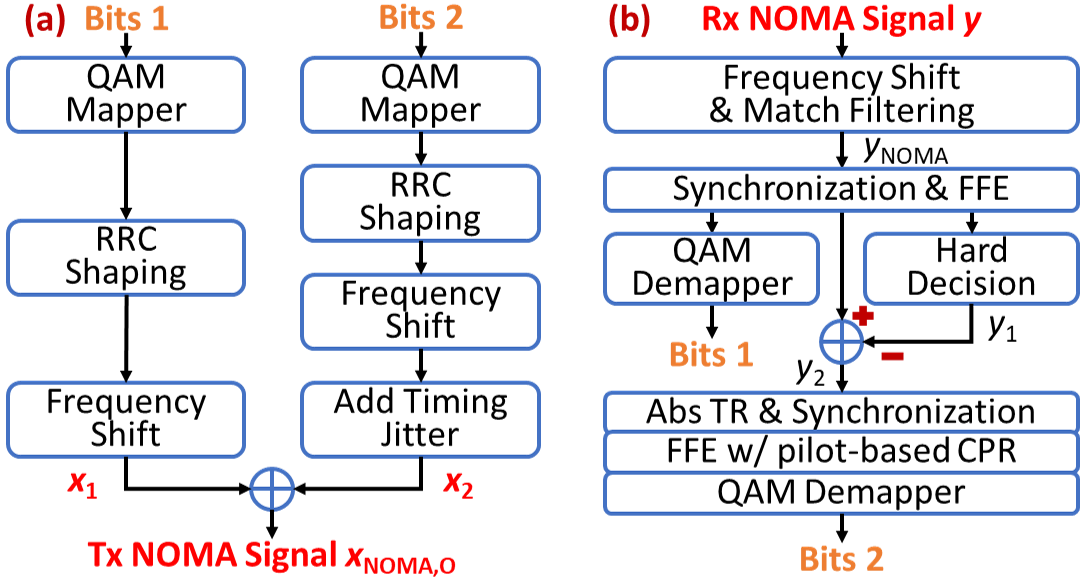}
\caption{Schematic diagram of (a) Tx DSP and (b) Rx DSP for NOMA with the asynchronous clocks.}
\label{Fig2}
\end{figure}

The frame structure and training sequence (TS) design for the NOMA signals is shown in Fig. \ref{Fig1}(a). One frame consists of TS and a payload sequence. The TS includes TS-A, TS-B, TS-C, and TS-D. The structure of TS-A and TS-B is shown in Fig. \ref{Fig1}(b). TS-A consisting of 16 sets of [0 0 1 1] sequences estimates the initial sampling phase for the TR algorithm. Frame synchronization is implemented based on TS-B. The sequence $\text{S}$ of TS-B consists of pseudo-random noise sequences with 20 symbols. Cyclic prefixes (CP) with a length of 4 are added to increase the tolerance for synchronization errors. TS-C is a training sequence of 500 quadrature phase shift keying (QPSK) symbols for equalizer training. TS-D is the pilot for the CPR \cite{zhou2024flexible, alfredsson2020pilot}. One pilot symbol is inserted into every 32 payload symbols to estimate the phase rotation of the received signal. 

\section{Experimental Setups}\label{Experimental_Setup}
The transmitter (Tx) digital signal processing (DSP) for NOMA with the asynchronous clocks is shown in Fig. \ref{Fig2}(a). The bit sequence 1 of strong ONU is mapped into the QPSK signal by a quadrature amplitude modulation (QAM) mapper. The signal is pulse-shaped by a raised root cosine (RRC) filter and frequency shifts to generate the signal $x_1(t)$. The Tx DSP for the weak ONU is the same as the strong ONU, while a timing jitter $\tau(t)$ is added to the signal $x_2(t)$ according to the ITU-T G.9807.1 to simulate the practical timing jitter caused by the asynchronous clocks \cite{ITUT10Gigabit}. After the electrical-to-optical conversion, the signal $x_1(t)$ and signal $x_2(t)$ are coupled in the optical domain to generate the Tx NOMA signal $x_{\text{NOMA,~O}}(t)$.

The receiver (Rx) DSP is shown in Fig. \ref{Fig2}(b). The received NOMA signal $y(t)$ is frequency-shifted and matched filtered to obtain the baseband signal $y_{\text{NOMA}}(t)$. After synchronization and equalization using a feed-forward equalizer (FFE), the bit sequence 1 is recovered by QAM de-mapping from the NOMA symbols. The strong ONU signal is regenerated by hard decisions and subtracted from the NOMA signal to obtain the regenerated weak ONU signal. The TR with Abs TED estimates and recovers the timing error. Then the frame synchronization eliminates the transmission delay $\Delta t$ between the two ONUs. The weak ONU signal is then equalized, and the phase noise is compensated using pilot-based CPR. Finally, the bit sequence 2 is recovered by QAM de-mapping. 

\begin{figure}[!t]
\centering
\includegraphics[width = \linewidth]{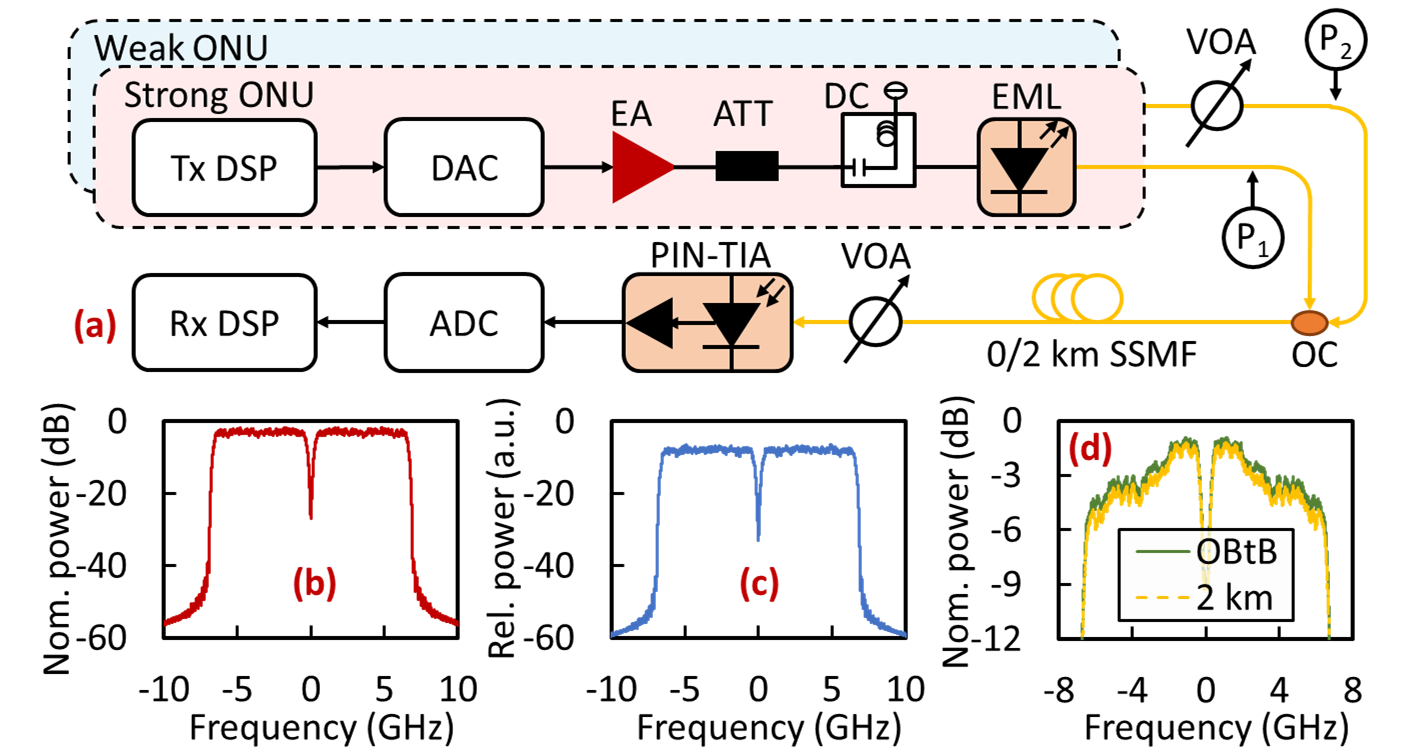}
\caption{(a) Experimental setups of the NOMA-PON with the asynchronous clocks. The spectrum of the transmitted signal at (b) strong and (c) weak ONUs. (d) The spectrum of received NOMA signal over OBtB and 2 km SSMF transmission at the OLT.}
\label{Fig3}
\end{figure}

Experimental setups of the NOMA-PON with the asynchronous clocks are shown in Fig. \ref{Fig3}(a). The 6.25 Gbaud digital signals $x_1(t)$ and $x_2(t)$ were uploaded into a digital-to-analog converter (DAC) with a sampling rate of 90 GSa/s and 3 dB bandwidth of 16 GHz. After being amplified by electrical amplifiers (EAs, Centellax OA4SMM4) followed by a 3 dB attenuator (ATT), the C-band electro-absorption modulated laser (EML) with a DC bias modulated the signals on the optical carriers to generate optical signals for strong and weak ONUs, respectively. The wavelengths of the two EMLs are $\sim1540.50$ nm and $\sim1543.00$ nm, respectively. The optical power of weak ONU was adjusted by a variable optical attenuator (VOA). The power ratio in dB is measured as $10 \log_{10} (P_{1} / P_{2})$, where $P_{1}$ and $P_{2}$ are the optical power of the strong and weak ONUs, respectively. Fig. \ref{Fig3}(b) and (c) show the transmitted signal spectrum at the power ratio of $5$ dB for strong and weak ONUs, respectively. The optical signals of the strong and weak ONUs were coupled into a NOMA signal by a 50:50 optical coupler (OC) and then fed into the 0/2 km standard single-mode fiber (SSMF), respectively. 

A VOA at the Rx was used to adjust the received optical power (ROP). The optical signal was converted into an electrical signal by a 40 GHz P-type-intrinsic-N-type diode with a trans-impedance amplifier (PIN-TIA, Finisar MPRV1331A). The analog signal was converted to a digital signal by an analog-to-digital converter (ADC) with a sampling rate of 90 GSa/s. Fig. \ref{Fig3}(d) shows the spectrum of the received signal over optical back-to-back (OBtB) and 2 km SSMF transmission, which shows that chromatic dispersion (CD) causes power fading. Since the C-band EMLs are used, a short-reach transmission is conducted due to CD. The NOMA-PON using C-band devices can be used for some short-reach access scenarios, while O-band devices or more advanced equalizers can be used to deal with the CD to extend the reach \cite{van2020strategies, sarmiento2020optical}. Finally, the Rx DSP shown in Fig. \ref{Fig2}(b) recovered the bits of strong and weak ONUs.

\begin{figure}[!t]
\centering
\includegraphics[width = 0.95\linewidth]{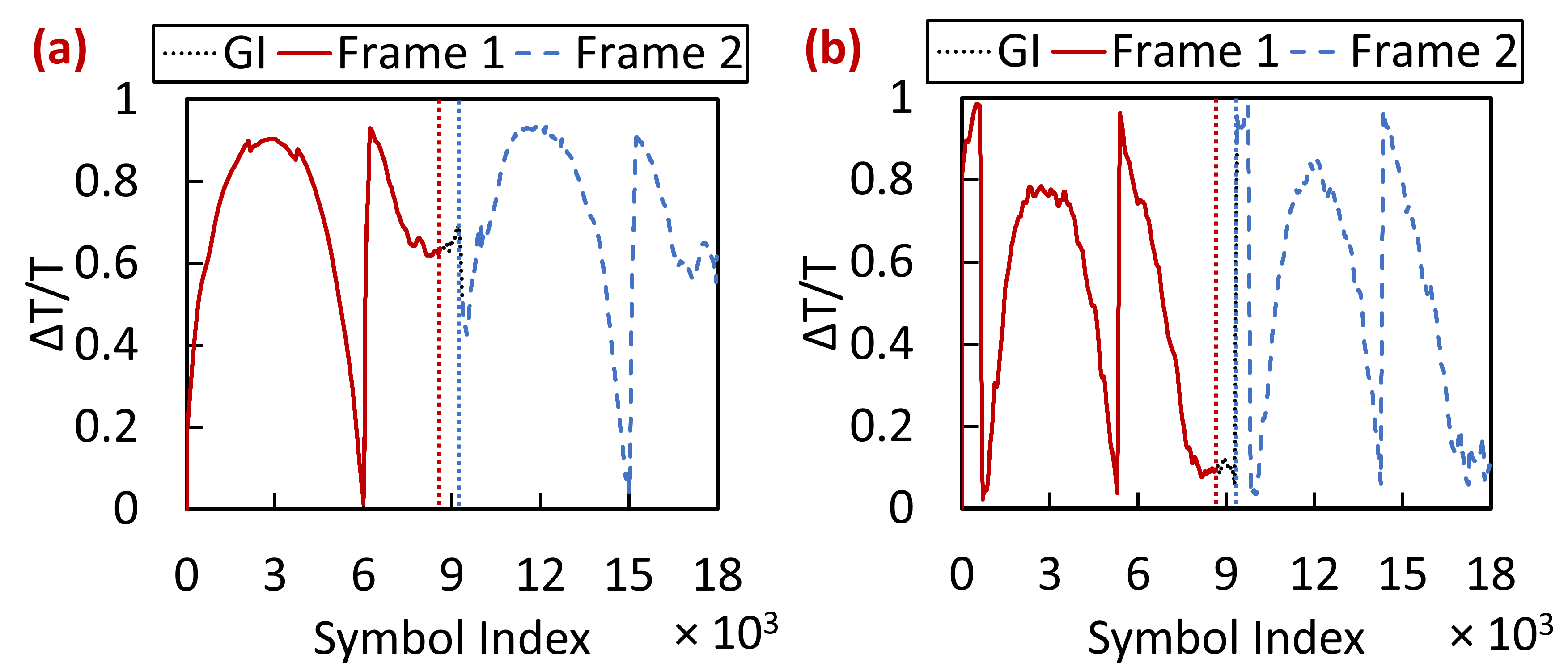}
\caption{Estimated timing error of weak ONU signals over (a) OBtB transmission and (b) 2 km SSMF transmission by using the TR with Abs TED. GI: guard interval.}
\label{Fig4}
\end{figure}

\begin{figure}[!t]
\centering
\includegraphics[width = 0.85\linewidth]{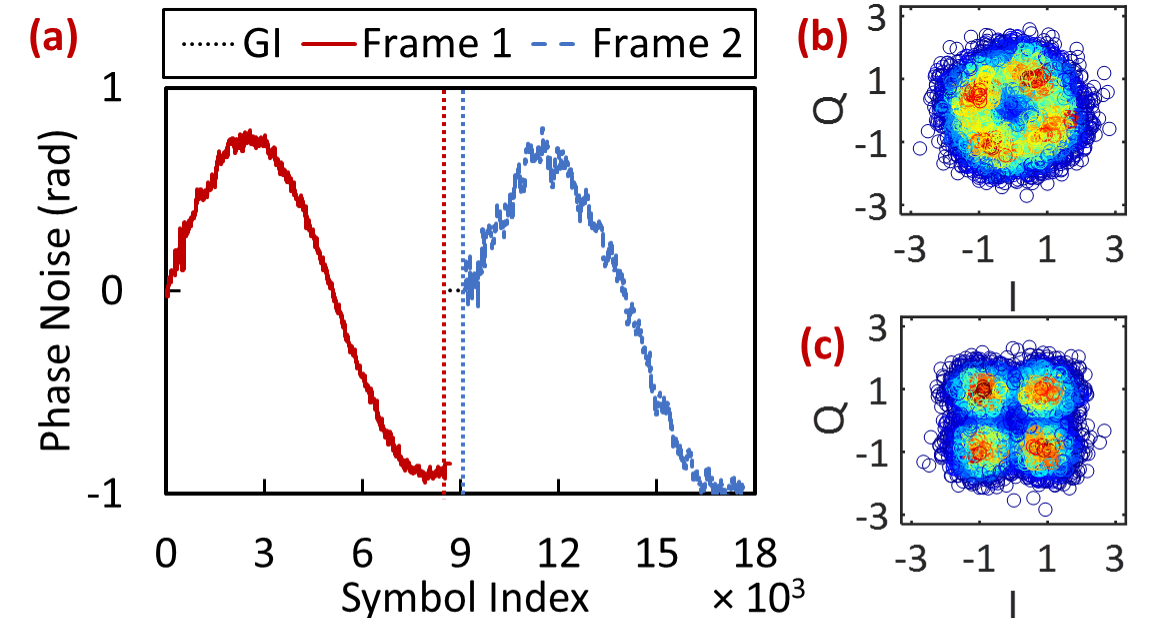}
\caption{(a) Estimated phase noise using the pilot-based CPR. The constellations of weak ONU (b) without and (c) with pilot-based CPR.}
\label{Fig5}
\end{figure}

\section{Experimental Results and Discussions}\label{Experimental_Results}
Figure \ref{Fig4} shows the estimated timing error of weak ONU signal over (a) OBtB transmission at the ROP of $-5$ dBm, and (b) 2 km SSMF transmission at the ROP of $-4$ dBm by using the TR with Abs TED. There are two frames and a guard interval (GI) between two frames. The duration of the GI is approximately $37.6$ ns. The TR with Abs TED at baud-rate sampling is used to estimate and compensate for the timing error, which is insensitive to phase noise by theoretical derivation. The estimated timing error approximates the $\tau(t)$ with the sin jitter model that is added at Tx, which proves that the TR with Abs TED can successfully track the clock and estimate the timing error in the presence of phase noise.

The estimated phase noise using the pilot-based CPR is shown in Fig. \ref{Fig5}(a). The GI between the two frames does not require CPR, which thus shows the blank in the estimated phase noise. The estimated phase noise also approximates the $\tau(t)$ with the sin model. The phase noise on the signal of the weak ONU leads to the phase rotation of the constellation point as shown in Fig. \ref{Fig5}(b). The constellation after compensating for the phase noise estimated using the pilot-based CPR is shown in Fig. \ref{Fig5}(c), where most of the constellation points rotated to the correct positions. Therefore, it proves that the pilot-based CPR can accurately estimate the phase noise caused by the asynchronous clocks.

\begin{figure}[!t]
\centering
\includegraphics[width = 0.9\linewidth]{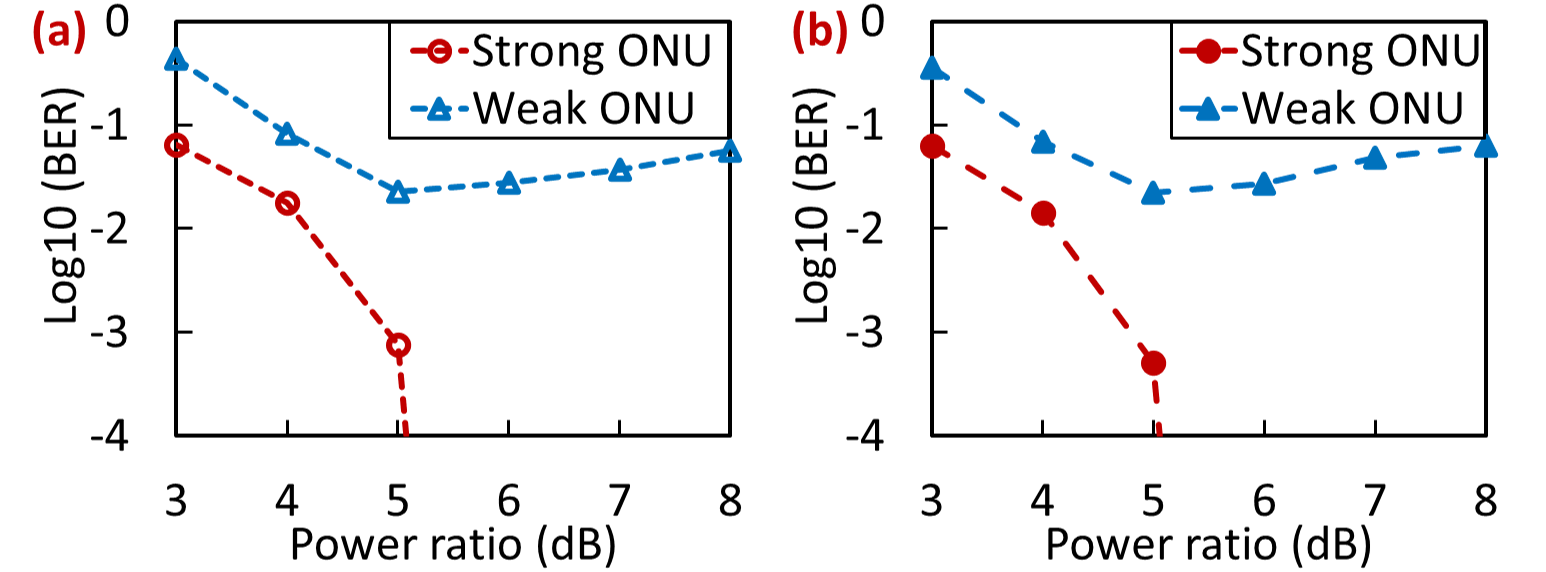}
\caption{BER versus power ratio for NOMA with the asynchronous clocks over (a) OBtB transmission and (b) 2 km SSMF transmission.}
\label{Fig6}
\end{figure}

Figure \ref{Fig6} shows the bit error ratio (BER) of strong and weak ONUs at different power ratios for NOMA-PON with the asynchronous clocks over (a) OBtB transmission at the ROP of $-6$ dBm and (b) 2 km SSMF transmission at the ROP of $-5$ dBm. As the power ratio decreases, the power of the weak ONU increases, and that of the strong ONU reduces, which leads to an increase and decrease in the BER of strong and weak ONUs, respectively. However, the signal of the weak ONU is recovered by subtracting the recovered signal of the strong ONU from the NOMA signal. When the power ratio is below $5$ dB, the BER of the weak ONU then increases due to the deterioration of the strong ONU signal. Thus, the power ratio is optimized at 5 dB to achieve the best BER performance of the weak ONU, while that of the strong ONU is still much better. The NOMA-PON with asynchronous clocks can be employed in access scenarios with low power differences.

\begin{figure}[!t]
\centering
\includegraphics[width = 0.9\linewidth]{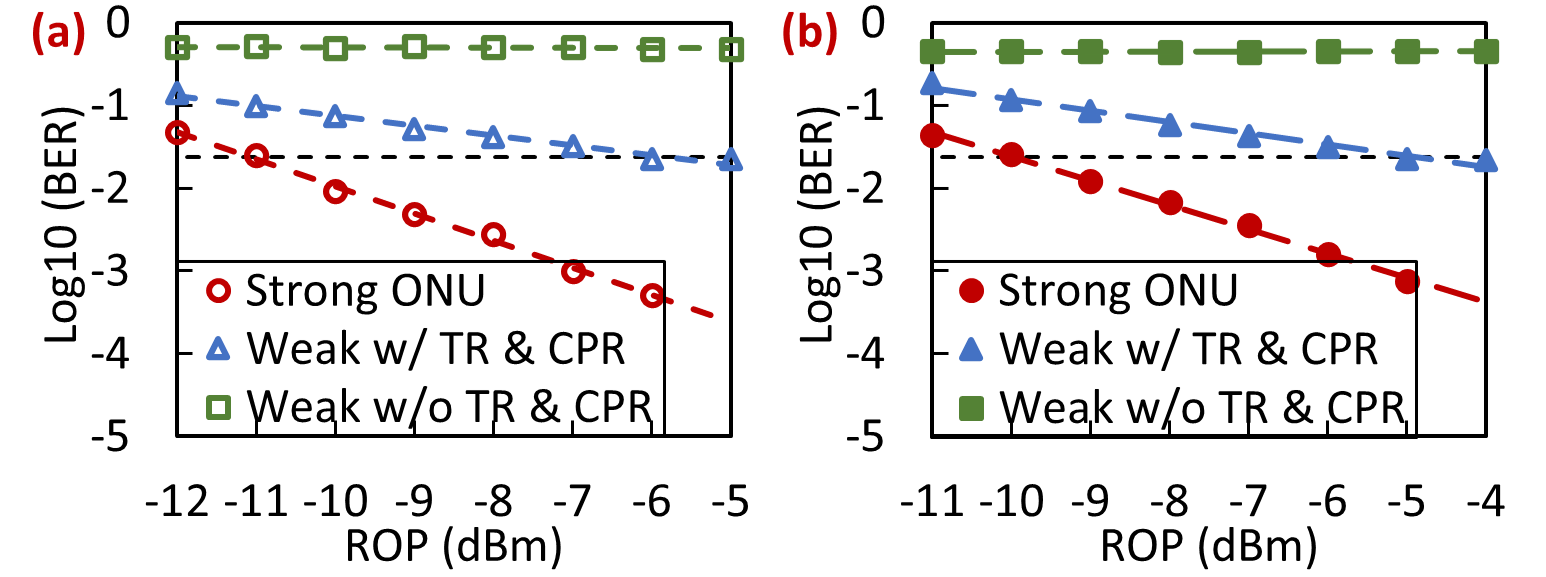}
\caption{BER versus ROP for NOMA-PON with the asynchronous clocks over (a) OBtB transmission and (b) 2 km SSMF transmission at the power ratio of $5$ dB. The black dashed line is the 20\% SD-FEC limit.}
\label{Fig7}
\end{figure}

Figure \ref{Fig7} shows the BER versus ROP for NOMA-PON with the asynchronous clocks over (a) OBtB and (b) 2 km SSMF transmission. BER of the weak ONU without the proposed TR algorithm and pilot-based CPR is about $0.5$ due to the timing error and phase noise. BER of the weak ONU over the OBtB and 2 km SSMF transmission using the proposed TR and pilot-based CPR can reach the 20\% soft-decision forward error correction (SD-FEC) limit at the ROP of $-5$ dBm and $-4$ dBm, respectively. Thus, the TR with Abs TED and the pilot-based CPR can effectively eliminate the timing error and phase noise for NOMA with asynchronous clocks. However, CD results in $\sim 1$ dB penalty in the receiver sensitivity for 2 km SSMF transmission compared to OBtB transmission. There are three possible reasons for the poor BER performance of weak ONU. Firstly, the 5-dB power ratio leads to the receiver sensitivity of the weak ONU being $\sim$5dB less than that of the strong ONU. Secondly, after the regenerated signal of the strong ONU is subtracted from the NOMA signal, the noise remains in the signal of the weak ONU and limits its BER performance. Thirdly, the timing jitter is added to the signal of weak ONU and causes phase noise, which also limits its BER performance.

\section{Conclusion}\label{Conclusion}
The asynchronous clocks between the strong and weak ONUs cause the timing error and phase noise on the signal of the weak ONU in the NOMA-PON. The theoretical derivation shows that the timing error and phase noise can be independently compensated. In this Letter, we propose the TR with Abs TED and pilot-based CPR to eliminate the timing error and phase noise. The experiment of 25G NOMA-PON is set up to verify the feasibility of the proposed algorithms. After compensating for timing error and phase noise, the weak ONU can reach the 20\% SD-FEC limit. In conclusion, the proposed TR and the pilot-based CPR show great potential for the NOMA-PON.

\bibliographystyle{IEEEtran}
\bibliography{IEEEexample}

\end{document}